\documentclass[twocolumn,showpacs,preprintnumbers,amsmath,amssymb]{revtex4}
\usepackage{amsmath,amssymb,graphics,epsfig,subfigure}
\usepackage{color}

\begin{document}
\thispagestyle{empty}

\begin{center}

\title{Insight into the Microscopic Structure of an AdS Black Hole from Thermodynamical Phase Transition}

\date{\today}
\author{Shao-Wen Wei\footnote{E-mail: weishw@lzu.cn} and
        Yu-Xiao Liu\footnote{Corresponding author. E-mail: liuyx@lzu.edu.cn}}

 \affiliation{Institute of Theoretical Physics, Lanzhou University,
           Lanzhou 730000, China}

\begin{abstract}
Comparing with an ordinary thermodynamic system, we investigate the possible microscopic structure of a charged anti-de Sitter black hole completely from the thermodynamic viewpoint. The number density of the black hole molecules is introduced to measure the microscopic degrees of freedom of the black hole. We found that the number density suffers a sudden change accompanied by a latent heat when the black hole system crosses the small-large black hole coexistence curve, while when the system passes the critical point, it encounters a second-order phase transition with a vanishing latent heat due to the continuous change of the number density. Moreover, the thermodynamic scalar curvature suggests that there is a weak attractive interaction between two black hole molecules. These phenomena might cast new insight into the underlying microscopic structure of a charged anti-de Sitter black hole.
\end{abstract}


\maketitle
\end{center}

%\section{Introduction}

{\emph{Introduction}.}---Black holes have been a mystery since they were predicted by general relativity. From a classical viewpoint, a black hole is a complete black object of strong gravity, and nothing can escape from it. However, after the pioneering work by Hawking and Bekenstein~\cite{Hawking,Bekensteina,Bekensteinb}, such a system was found to possess temperature $T$ and entropy $S$, i.e.,
\begin{equation}
 T=\frac{\hbar\kappa}{2\pi ck_{B}},\quad S=\frac{k_{B}c^{3}A}{4\hbar G},
\end{equation}
with $\kappa$ and $A$ the surface gravity and event horizon area of the black hole. For simplicity, we adopt in the following the units $\hbar=c=k_{B}=G=1$. Then the gravity system is mapped to a thermodynamic system.
In Ref.~\cite{Bardeen}, four laws of black hole thermodynamics were established. Later, many works show that a black hole is not only a gravity system, but also a special thermodynamic system. Comparing with an ordinary thermodynamic system, understanding the microscopic origin of black hole entropy is a challenging problem, because the black hole entropy is proportional to the horizon area, i.e., $S\sim A$ rather than the volume. Such a subject attracts a great interest, especially on the microscopic degrees of freedom of a black hole, which, however, is still not completely clear.

Among the methods of counting black hole microstates and investigating the
microscopic origin of black hole entropy, string theory provides a natural framework. Through counting the number of states of a weakly coupled D-brane system and then extrapolating the result to the black hole phase, the Bekenstein-Hawking entropy formula was exactly derived for certain supersymmetric black holes by Strominger and Vafa~\cite{Vafa}. Similar calculations have been applied to other black holes~\cite{Maldacena,Callan,Horowitz,Emparan}. Despite the great success, such calculation is limited in supersymmetric and extremal black holes. For example, the explicit construction of the microstates for the most simple Schwarzschild and Kerr black hole solutions is still lacking. There are other ways to calculate black hole entropy, but almost all of them are based on the assumption that gravity is dual to a gauge theory or a strongly coupled conformal field theory, and black hole entropy and microstates are calculated under this duality. Nevertheless, what constitutes the states of a black hole is unclear. Furthermore, fuzzball theory \cite{Lunin,Mathur} states that black holes are actually spheres of strings with a definite volume from the microscopic level. However, it is still highly speculative.

Maybe the unsatisfactory understanding of the microscopic structure of a black hole brings us back to the old question: Does a black hole have a microscopic structure? For the fluid, its microscopic structure is very clear, i.e., its micromolecules carry the degree of freedom. How about a black hole? Its microscopic structure is completely unknown to us. In order to answer this question, we recall Boltzmann's insight: ``If you can heat it, it has microscopic structure". Such a viewpoint sheds insight into the microscopic structure of matter before achieving observational evidence in the past. Since a black hole can change its Hawking temperature by absorbing or emitting matter, we can conjecture that it should have a microscopic structure, even though we do not know its micromolecules. The aim of this Letter is to explore the possible microscopic structure of a charged anti-de Sitter (AdS) black hole completely from the thermodynamic viewpoint.

%\section{Number density of the microscopic structure}

{\emph{Number density of the microscopic structure}.}---Next we will examine the number density of the black hole microscopic structure from the point of view of the thermodynamic phase transition. A powerful model to describe a phase transition is the van der Waals (vdW) fluid, which is the first, simplest, and most widely known example of an interacting thermodynamic system exhibiting a first-order liquid-gas phase transition. About sixteen years ago, it was found that the phase transition between small and large charged AdS black holes is of the vdW type in the canonical ensemble~\cite{Chamblin,Chamblin2}. However, until recently, the complete analogy between the vdW fluid and charged AdS black hole was established by treating the cosmological constant as a pressure, $P=-\Lambda/8\pi$~\cite{Dolan,Kubiznak}. The first step of analogy between the vdW fluid and black hole system is the equation of state. For a four-dimensional charged AdS black hole, it reads~\cite{Kubiznak}
\begin{equation}
 P=\frac{T}{2r_{h}}-\frac{1}{2\pi (2r_{h})^{2}}+\frac{2Q^{2}}{\pi (2r_{h})^{4}}.\label{stateequation}
\end{equation}
Here $r_{h}$, $Q$ and $T$ are the horizon radius, temperature, and charge of the black hole, respectively. Comparing with the vdW fluid, the specific volume $v$ of the \emph{black hole fluid} can be identified as~\cite{Kubiznak}
\begin{equation}
 v=2l_{P}^{2}r_{h},
\end{equation}
where we restore the dimension and Planck length $l_{P}=\sqrt{\hbar G/c^{3}}$. This concept has been applied to different black hole systems and is a great success in studying the small/large black hole (SBH/LBH) phase transition. Thus, we can understand $v$ as the specific volume of the \emph{black hole molecule}, which carries the degrees of freedom of black hole entropy.

On the side of the vdW fluid, the number density of micromolecules is a fundamental physical quantity to describe the thermodynamic system running in the phase diagram. And for the black hole fluid, we can introduce the concept, which is defined as
\begin{equation}
 n=\frac{1}{v}=\frac{1}{2l_{P}^{2}r_{h}}.\label{NN}
\end{equation}
We will show that this concept could provide a preliminary knowledge on the microscopic structure of a black hole. With such a quantity $n$, the microscopic and macroscopic physical quantities of a thermodynamic system are closely related to each other through statistical mechanics. And novel interesting information will be revealed.

However, before pursuing this issue, we will turn back to the number density $n$.  Its introduction seems to be somewhat incredible. In what follows we will give a natural interpretation for it.

From the holographic view, black hole entropy resides on the black hole horizon. Ruppeiner \cite{Ruppeiner0} proposed that the microscopic degrees of freedom of the black hole are carried by the Planck area pixels, i.e., $A/l_{P}^{2}$. Here, we  assume that one microscopic degree of freedom occupies $\gamma$ Planck area pixels. Then the total number of the microscopic degrees of freedom is given by
\begin{equation}
 N=\frac{A}{{\gamma}l_{P}^{2}}.
\end{equation}
Therefore, the effective number density $n$ for the AdS black hole can be calculated as \cite{Altamirano2}
\begin{equation}
 n=\frac{N}{V}=\frac{3}{{\gamma}l_{P}^{2} r_{h}},
\end{equation}
where $V=\frac{4{\pi}r_{h}^{3}}{3}$ is the thermodynamic volume produced from the first law, $V=(\partial_{P}M)_{Q,S}$, rather than the volume of the sphere with the radius of the event horizon~\cite{Dolan,Kastor}. Taking ${\gamma}=6$, we will obtain the exact result Eq. (\ref{NN}). For a detailed discussion on the black hole microscopic degree of freedom, we refer the readers to Refs.~\cite{Padmanabhan0,Banerjee,Padmanabhan}.

Here we grasp that the number density $n$ defined here has a natural interpretation, and it can measure the number density of the virtual black hole molecules. Finding the relations between the thermodynamic quantities and number density can also provide an effective link between microscopic and macroscopic black hole physics.

%\section{Thermodynamic phase transition}

{\emph{Thermodynamic phase transition}.}---The state equation (\ref{stateequation}) of a charged AdS black hole displays a vdW-like thermodynamic behavior. The SBH-LBH coexistence curve has a parametric form~\cite{Wei2014}
\begin{equation}
 \frac{P}{P_{c}}=\sum_{i} a_{i}\Big(\frac{T}{T_{c}}\Big)^{i}, \label{coex4}
\end{equation}
with $a_{i}$ are dimensionless coefficients. This curve has a positive slope everywhere and terminates at the critical point $(P_{c}, T_{c}, v_{c})=(1/96\pi Q^{2}, \sqrt{6}/18\pi Q, 2\sqrt{6}Q)$, beyond which the SBH and LBH phases cannot be clearly distinguished. Near this critical point, the charged AdS black hole system shares the same critical exponents and scal  ing laws with the vdW fluid. This strongly suggests that such a SBH/LBH phase transition is of the vdW type, and now it is widely accepted.

%%%%%%%%%%%%%%%%%%%%%%%%%%%%%%%%%%%%%%%%%%%%%%%%%%%%%%%%%%%%%%%%%%%%%
\begin{figure}
\includegraphics[width=7cm]{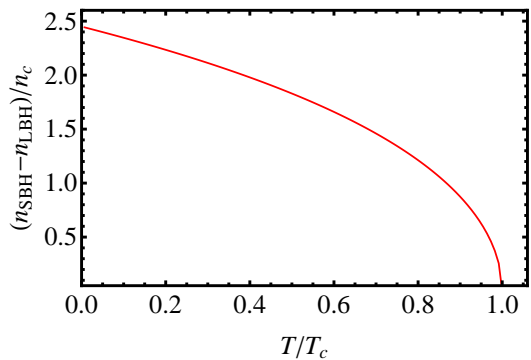}
\caption{The difference of the number densities between the small and large black holes.}\label{pnt}
\end{figure}
%%%%%%%%%%%%%%%%%%%%%%%%%%%%%%%%%%%%%%%%%%%%%%%%%%%%%%%%%%%%%%%%%%%%%%%%%

Recently there have been many groups finding the phase transition of the vdW type in different AdS black hole backgrounds. But there are few papers considering the physics going along or crossing the SBH-LBH coexistence curve. Studying such an issue will raise new interest in black hole thermodynamics.

\begin{figure}
\includegraphics[width=7cm]{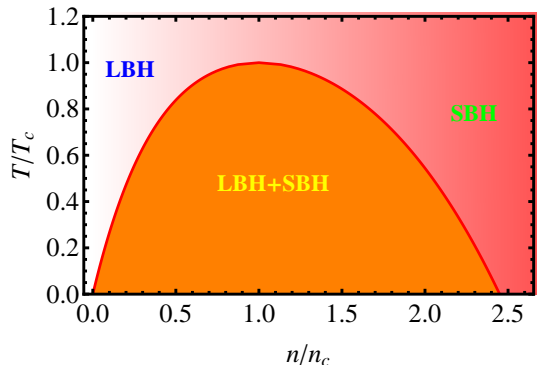}
\caption{Phase diagram in the $T-n$ plane. The black hole system has a maximum number density at $n/n_{c}=2.44$, slightly smaller than $n/n_{c}=3$ of the van der Waals fluid.}\label{pttnn}
\end{figure}

When a SBH crosses the coexistence curve and becomes a LBH, the number density would have a discontinuous change. We show the behavior of the difference of the number density between the SBH and LBH along the curve in Fig.~\ref{pnt}, which displays that with an increase of the temperature, the difference $(n_{SBH}-n_{LBH})/n_{c}$ monotonically decreases. It vanishes when approaching the critical point, which implies that the microscopic structures of the SBH and LBH tend to be the same at that point. The temperature $T/T_{c}$ versus density $n/n_{c}$ phase diagram is plotted in Fig.~\ref{pttnn}. The maximum number density of the system is $n/n_{c}=2.44$, slightly smaller than $n/n_{c}=3$ of the vdW fluid~\cite{Johnston}, at which there is no free volume left for molecules to move. The experimental data of $Ne$, $Ar$, $O_{2}$ and other gases approximatively confirmed such a phase diagram. Indeed, we can conjecture that the similar phenomena also holds for the black hole.

%%%%%%%%%%%%%%%%%%%%%%%%%%%%%%%%%%%%%%%%%%%%%%%%%%%%%%%%%%%%%%%%%%%%%
%%%%%%%%%%%%%%%%%%%%%%%%%%%%%%%%%%%%%%%%%%%%%%%%%%%%%%%%%%%%%%%%%%%%%%%%%

%%%%%%%%%%%%%%%%%%%%%%%%%%%%%%%%%%%%%%%%%%%%%%%%%%%%%%%%%%%%%%%%%%%%%
\begin{figure}
\includegraphics[width=8cm]{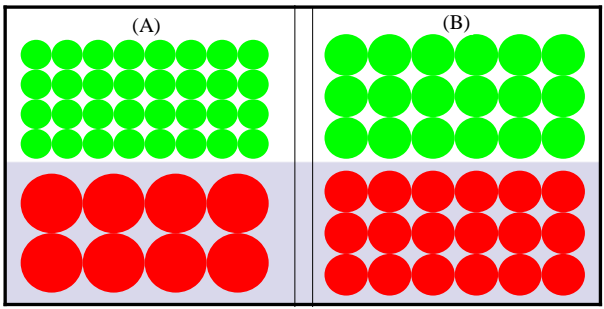}
\caption{Sketch picture for the change of the black hole molecule specific volume crossing the SBH-LBH coexistence curve. Top green and bottom red circles denote the small and large black hole molecules, respectively. During the phase transition, the specific volume of the molecular and number density change accordingly. (A) $(P, T)=(0.85 P_{c}, 0.94 T_{c})$, $v=0.73v_{c}$ (top), $1.48 v_{c}$ (bottom). (B) $(P, T, v)=(P_{c}, T_{c}, v_{c})$.}\label{pslbh}
\end{figure}
%%%%%%%%%%%%%%%%%%%%%%%%%%%%%%%%%%%%%%%%%%%%%%%%%%%%%%%%%%%%%%%%%%%%%%%%%

We show in Fig.~\ref{pslbh} the sketch picture showing that phase transition occurs between the small and large black hole molecules when the back hole system crosses the coexistence curve. The green circles in the upper half plane correspond to the SBH molecules, and the red circles in the lower half plane to the LBH molecules. Plane (A) denotes the situation that the black hole crosses the coexistence curve with $(P, T, v)<(P_{c}, T_{c}, v_{c})$, and (B) for $(P, T, v)=(P_{c}, T_{c}, v_{c})$. From Fig.~\ref{pslbh}(A), we see that when the black hole crosses the coexistence curve, the number density $n$ and the specific volume $v$ suffer a gap corresponding to a first-order phase transition. However, when the black hole passes the critical point, see Fig.~\ref{pslbh}(B), they continuously change, which corresponds to a second-order phase transition.

Generally, if the microscopic structure of an ordinary thermodynamic system has a discontinuous change during the phase transition, then there must be a nonvanishing latent heat. For the case of a black hole with fixed charge, the latent heat $L$ of each black hole molecule transiting from one phase to another phase can be calculated from the following formula:
\begin{equation}
 L=\frac{T\Delta S}{N}=T\Delta v\frac{dP}{dT}=T  \left(\frac{1}{n_{\text{LBH}}}-\frac{1}{n_{\text{SBH}}}\right) \frac{dP}{dT},
\end{equation}
where the Clapeyron equation $dP/dT=\Delta S/\Delta V$ holding along the coexistence curve has been used. From this equation, one can see that, when the black hole system crosses the coexistence curve, there is a nonvanishing latent heat, while the latent heat vanishes when the system passes the critical point $T=T_{c}$ due to $n_{\text{LBH}}=n_{\text{SBH}}$, which can be found from Fig.~\ref{pnt} or Fig.~\ref{pslbh}(B).

%\section{Thermodynamic geometry}
{\emph{Thermodynamic geometry}.}---Now it is clear that the number density $n$ measuring the micromolecules of black hole freedom is a useful quantity to describe the SBH/LBH phase transition. For the vdW fluid, interactions between two molecules are approximated by
the so-called Lennard-Jones potential. It produces short-range repulsive interaction and longer-range attractive one. Analogous to this, one would like to ask what kind of interaction there is between two micromolecules of a black hole. In order to answer this question, exploration of the thermodynamic fluctuation theory is necessary. Fortunately, the well-known thermodynamic geometry, the Ruppeiner geometry~\cite{Ruppeiner}, constructed from the thermodynamic fluctuation theory, provides us a powerful tool.

The line element of the Ruppeiner geometry is defined in parameter space by taking the system entropy $S$ as its thermodynamic potential,
\begin{equation}
 ds^{2}=\frac{\partial^{2}S}{\partial x^{\mu}\partial x^{\nu}}
     \Delta x^{\mu}\Delta x^{\nu},\label{metric}
\end{equation}
where $\Delta x^{\mu}\equiv(x^{\mu}-x^{\mu}_{0})$ measures the fluctuation of the thermodynamic quantity $x^{\mu}$ from $x^{\mu}_{0}$, and $x^{\mu}_{0}$ corresponds to $ds^{2}=0$. Such a line element has a clear physical interpretation~\cite{Ruppeiner}: the less probable a fluctuation between two thermodynamic states, the further apart they are.

%%%%%%%%%%%%%%%%%%%%%%%%%%%%%%%%%%%%%%%%%%%%%%%%%%%%%%%%%%%%%%%%%%%%%
\begin{figure}
\includegraphics[width=7cm]{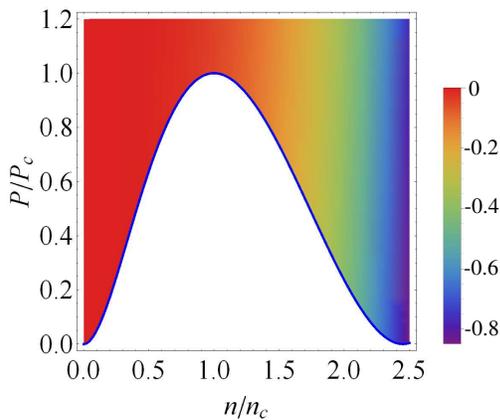}
\caption{The Ruppeiner scalar curvature $R$ as a function of the pressure $P$ and number density $n$. The boundary marked as the thick blue line is related to the coexistence curve.}\label{pr}
\end{figure}
%%%%%%%%%%%%%%%%%%%%%%%%%%%%%%%%%%%%%%%%%%%%%%%%%%%%%%%%%%%%%%%%%%%%%%%%%

Based on the metric (\ref{metric}), one can construct a thermodynamic scalar curvature $R$ similar to that of GR. The authors in Ref.~\cite{Oshima} first emphasized the difference in the sign of $R$, i.e., $R>0$ ($R<0$) for the Fermi (Bose) ideal gas and $R=0$ for the classical ideal gas \cite{Rsign}. Further study shows that positive (negative) $R$ implies repulsive (attractive) interaction dominates in the thermodynamic system. Thus the sign of $R$ offers us direct information about the character of the interaction among the micromolecules of a thermodynamic system.

For the charged AdS black hole, we take $(M, P)$ fluctuation while with fixed charge $Q=1$. This case corresponds to a two-dimensional Riemannian geometry with the curvature scalar given by
\begin{equation}
 R=-\frac{4(n/n_{c})^{4}+(n/n_{c})^{6}}{24\pi (n/n_{c})^{2}+12\pi (P/P_{c})}.\label{curvature}
\end{equation}
The behavior of $R$ is depicted in Fig.~\ref{pr}.
Note that in the range of $(n/n_{c}, P/P_{c})<(1, 1)$, small and large $n$ relate to the LBH and SBH, respectively. Between the small and large black holes, it is the intermediate black hole, which has been excluded from thermodynamic phase transition consideration for its instability. Thus, the curvature scalar $R$ has a gap among the SBH and LBH branches. This incommensurate $R$~\cite{Ruppeiner3} implies that the structures of the SBH and LBH are different, and it is difficult for the black hole molecules to transform from one phase to another unless providing the latent heat. At the asymptotic critical region $(n/n_{c}, P/P_{c})\sim(1, 1)$, the gap of $R$ in the coexisting SBH and LBH phases disappears. In such case, the structures of the SBH and LBH are similar. Phase transition between them can easily take place and the latent heat in the process tends to vanish, which is in accord with our above analysis. With further increase of the parameters such that $(n/n_{c}, P/P_{c})>(1, 1)$, $R$ continuously varies. This is due to the fade out of the intermediate black hole branch, and no clear boundary between the SBH and LBH branches.

Since $(n/n_{c})$ and $(P/P_{c})$ are always positive, $R$ is negative according to Eq. (\ref{curvature}). From thermodynamic fluctuation analysis~\cite{Ruppeiner2}, this case is related to the system of a weak attractive interaction between two black hole molecules. In a fixed AdS space, the very large black hole has $n/n_{c}\sim 0$, and thus the curvature scalar $R$ approaches zero. This reveals that the property of the very large black hole microscopic structure is similar to that of the classical ideal gas, and there exists no interaction between the micromolecules. While for the small black hole, the attractive interaction grows stronger with $(n/n_{c})^{4}$. Such patterns of $R$ might cast new sight on the nature of the AdS black hole microscopic properties.

%\section{Summary}
{\emph{Summary}.}---Before ending this Letter, we briefly summarize it. We tried to explore the microscopic structure of a charged AdS black hole from the viewpoint of thermodynamic phase transition and thermodynamic geometry. After making a comparison with the vdW fluid, we introduced a new concept, the number density $n$, for the black hole with the assumption that it has a microscopic structure. Employing it, we studied the behaviors of the macroscopic thermodynamic variables. The result shows that when the system crosses the SBH-LBH coexistence curve, the specific volume of the black hole molecule suffers a sudden change, accompanied by a latent heat. However, when the system crosses the critical point, the latent heat vanishes, which implies a second-order phase transition. This is due to the fact that the specific volumes of the small and large black hole molecules tend to be the same when approaching that point. On the other hand, we found that the interaction between two black hole molecules is weak attractive from the viewpoint of the thermodynamic geometry. These results might cast new insight into the black hole microscopic structure. And it seems that the thermodynamic phase transition and thermodynamic geometry could offer us an effective method to study the microscopic properties of a black hole.

In this Letter, we showed two important quantities, the number density $n$ and specific volume $v$, to describe the black hole molecule. However, what the black hole microstates actually are is still unknown. Maybe the microscopic degrees of freedom carried by these molecules can be counted by the D-brane states. Or, more speculatively, one can think that a molecule is a sphere of strings following the fuzzball proposal. Thus, more effort is still needed in order to understand the real microscopic structure of a black hole from a much more basic viewpoint in the future.

Last, we would like to make a few comments. Here we only deal with the charged AdS black hole. How do non-AdS black holes behave? We conjecture that these black holes also have a microscopic structure from Boltzmann's insight. However, thermodynamic phase transition of the vdW type may not exist and thus the system behaves like the classical ideal gas. Furthermore, one can also consider rotating AdS black holes or black holes in higher-derivative gravity~\cite{Altamirano1,Altamirano2,Frassino2014,Wei}, where richer phase transition structure has been found and it may provide an extensive insight into the black hole microscopic structure.

%\section*{Acknowledgements}
{\emph{Acknowledgements}.}---We would like to thank the anonymous referees whose suggestions and comments largely helped us in improving the original manuscript. We also thank Professor Li-Ming Cao and Dr. Hai-Shan Liu for useful discussion. This work was supported by the National Natural Science Foundation of China (Grants No. 11205074 and No. 11375075).


\begin{thebibliography}{99}

\bibitem{Hawking}
 S. W. Hawking,
% {\em Particle creation by black holes},
   Commun. Math. Phys. \textbf{43}, 199 (1975).

\bibitem{Bekensteina}
 J. Bekenstein,
%   {\em Black holes and the second law},
     Lett. Nuovo Cim. \textbf{4}, 737 (1972).

\bibitem{Bekensteinb}
 J. D. Bekenstein,
%  {\em Black holes and entropy},
    Phys. Rev. \textbf{D 7}, 2333 (1973).

\bibitem{Bardeen}
  J. M. Bardeen, B. Carter, and S. Hawking,
%   {\em The four laws of black hole mechanics},
    Commun. Math. Phys.\textbf{31}, 161 (1973).

\bibitem{Vafa}
  A. Strominger and C. Vafa,
%   {\em Microscopic origin of the Bekenstein-Hawking entropy},
     Phys. Lett. \textbf{B 379}, 99 (1996).
    %arXiv:hep-th/9601029

\bibitem{Maldacena}
 J. M. Maldacena and A. Strominger,
%   {\em Statistical entropy of four-dimensional extremal black holes},
      Phys. Rev. Lett. \textbf{77}, 428 (1996).

\bibitem{Callan}
 C. G. Callan and J. M. Maldacena,
%  {\em D-brane approach to black hole quantum mechanics},
    Nucl. Phys. \textbf{B 472}, 591 (1996).

\bibitem{Horowitz}
   G. T. Horowitz and A. Strominger,
%    {\em Counting states of near extremal black holes},
     Phys. Rev. Lett. \textbf{77}, 2368 (1996).

\bibitem{Emparan}
   R. Emparan and G. T. Horowitz,
%    {\em Microstates of a neutral black hole in M theory},
      Phys. Rev. Lett. \textbf{97}, 141601 (2006).


\bibitem{Lunin}
   O. Lunin, S. D. Mathur,
%    {\em AdS/CFT duality and the black hole information paradox},
      Nucl. Phys. \textbf{B 623}, 342 (2002).

\bibitem{Mathur}
   O. Lunin, S. D. Mathur,
%    {\em Statistical interpretation of Bekenstein entropy for systems with a stretched horizon},
      Phys. Rev. Lett. \textbf{88}, 211303 (2002).

\bibitem{Chamblin}
 A. Chamblin, R. Emparan, C. V. Johnson, and R. C. Myers,
%  {\em Charged AdS black holes and catastrophic holography},
   Phys. Rev. \textbf{D 60}, 064018 (1999), [arXiv:hep-th/9902170].


\bibitem{Chamblin2}
 A. Chamblin, R. Emparan, C. V. Johnson, and R. C. Myers,
%  {\em Holography, Thermodynamics and fluctuations of charged AdS black holes},
   Phys. Rev. \textbf{D 60}, 104026 (1999), [arXiv:hep-th/9904197].

\bibitem{Dolan}
  B. P. Dolan,
%   {\em The cosmological constant and the black hole equation of state},
  Class. Quant. Grav. \textbf{28}, 125020 (2011), [arXiv:1008.5023[gr-qc]].

\bibitem{Kubiznak}
   D. Kubiznak and R. B. Mann,
%   {\em P-V criticality of charged AdS black holes},
     JHEP \textbf{1207}, 033 (2012), [arXiv:1205.0559[hep-th]].


\bibitem{Ruppeiner0}
   G. Ruppeiner,
%   {\em Thermodynamic curvature and phase transitions in Kerr-Newman black holes},
     Phys. Rev. \textbf{D 78}, 024016 (2008), [arXiv:0802.1326[gr-qc]].

\bibitem{Altamirano2}
 N. Altamirano, D. Kubiznak, R. B. Mann, and Z. Sherkatghanad,
%    {\em Thermodynamics of rotating black holes and black rings: phase transitions and thermodynamic volume },
     Galaxies \textbf{2}, 89 (2014) [arXiv:1401.2586[hep-th]].

\bibitem{Kastor}
  D. Kastor, S. Ray, and J. Traschen,
%   {\em Enthalpy and the mechanics of AdS black holes},
   Class. Quant. Grav. \textbf{26}, 195011 (2009), [arXiv:0904.2765[hep-th]].


\bibitem{Padmanabhan0}
  T. Padmanabhan,
%   {\em Entropy of static spacetimes and microscopic density of states},
  Class. Quant. Grav. \textbf{21}, 4485 (2004), [arXiv:gr-qc/0308070].


\bibitem{Banerjee}
  R. Banerjee and B. R. Majhi,
%   {\em Statistical Origin of Gravity},
  Phys. Rev. \textbf{D 81}, 124006 (2010), [arXiv:1003.2312[gr-qc]].

\bibitem{Padmanabhan}
  T. Padmanabhan,
%   {\em Surface density of spacetime degrees of freedom from equipartition law in theories of gravity},
  Phys. Rev. \textbf{D 81}, 124040 (2010), [arXiv:1003.5665[gr-qc]].


\bibitem{Wei2014}
  S.-W. Wei and Y.-X. Liu,
   %{\em Clapeyron equations and fitting formula of the coexistence curve
   %in the extended phase space of the charged AdS black holes},
   Phys. Rev. \textbf{D 91}, 044018 (2015),
  [arXiv:1411.5749[hep-th]].

\bibitem{Johnston}
  D. C. Johnston,
   {\em Thermodynamic properties of the van der Waals fluid},
  [arXiv:1402.1205[cond-mat.soft]].

\bibitem{Ruppeiner}
 G. Ruppeiner,
%  {\em Riemannian geometry in thermodynamic fluctuation theory},
    Rev. Mod. Phys. \textbf{67}, 605 (1995);
    erratum ibid \textbf{68}, 313 (1996).

\bibitem{Oshima}
 H. Oshima, T. Obata, and H. Hara,
%  {\em Riemann scalar curvature of ideal quantum gases obeying Gentile's statistics},
   J. Phys. A: Math. Gen. \textbf{32}, 6373 (1999).

\bibitem{Rsign}
Here we adopt the same sign convention for $R$ with \cite{Ruppeiner}.

\bibitem{Ruppeiner3}
 G. Ruppeiner,
%  {\em Thermodynamic curvature from the critical point to the triple point},
     Phys. Rev. \textbf{E 86}, 021130 (2012), [arXiv:1208.3265[cond-mat.stat-mech]].

\bibitem{Ruppeiner2}
 G. Ruppeiner,
%  {\em Thermodynamic curvature and black holes},
    Springer Proc. Phys. \textbf{153}, 179 (2014), [arXiv:1309.0901[gr-qc]].

\bibitem{Altamirano1}
 N. Altamirano, D. Kubiznak, and R. B. Mann,
%  {\em Reentrant phase transitions in rotating AdS black holes},
   Phys. Rev. \textbf{D 88}, 101502 (2013), [arXiv:1306.5756[hep-th]].

\bibitem{Frassino2014}
 A. M. Frassino, D. Kubiznak, R. B. Mann, and F. Simovic,
%     {\em Multiple reentrant phase transitions and triple points in Lovelock thermodynamics},
       JHEP \textbf{1409}, 080 (2014), [arXiv:1406.7015[hep-th]].

\bibitem{Wei}
  S.-W. Wei and Y.-X. Liu,
  %   {\em Triple points and phase diagrams in the extended phase space of
   %charged Gauss-Bonnet black holes in AdS space},
  Phys. Rev. \textbf{D 90}, 044057 (2014),
  [arXiv:1402.2837[hep-th]].



\end{thebibliography}
\end{document}